\documentclass{emulateapj}

\def\kms    {\ifmmode{{\rm \ts km\ts s}^{-1}}\else{\ts km\ts s$^{-1}$}\fi}
\def\msol   {\ifmmode{{\rm M}_{\odot}}\else{M$_{\odot}$}\fi}
\def\ts     {\thinspace} 
\def\ci   {\ifmmode{{\rm C}{\rm \small I}}\else{C\ts {\scriptsize I}}\fi}
\def\cii  {\ifmmode{{\rm C}{\rm \small II}}\else{C\ts {\scriptsize II}}\fi}
\def\hi   {\ifmmode{{\rm H}{\rm \small I}}\else{H\ts {\scriptsize I}}\fi}
\def\hii  {\ifmmode{{\rm H}{\rm \small II}}\else{H\ts {\scriptsize II}}\fi}

\shorttitle{Cold Dust in the Tidal \hi\  Arms of the M\,81 Triplet}

\shortauthors{Walter et al.}

\journalinfo{}
\begin{document}

\title{The Displaced Dusty ISM of NGC\,3077: Tidal Stripping in the M\,81 Triplet.}
\author{
F. Walter\altaffilmark{1}
K. Sandstrom\altaffilmark{1} 
G. Aniano\altaffilmark{2}
D. Calzetti\altaffilmark{3}
K. Croxall\altaffilmark{4}
D.~A. Dale\altaffilmark{5}
B.~T. Draine\altaffilmark{2}
C. Engelbracht\altaffilmark{6}
J. Hinz\altaffilmark{6}
R.~C. Kennicutt\altaffilmark{7}
M. Wolfire\altaffilmark{8}
L. Armus\altaffilmark{9}
P. Beir\~{a}o\altaffilmark{9}
A.~D. Bolatto\altaffilmark{8}
B. Brandl\altaffilmark{10} 
A. Crocker\altaffilmark{3}
M. Galametz\altaffilmark{7}
B. Groves\altaffilmark{10}
C.-N. Hao\altaffilmark{11}
G. Helou\altaffilmark{12}
L. Hunt\altaffilmark{13}
J. Koda\altaffilmark{14}
O. Krause\altaffilmark{1} 
A. Leroy\altaffilmark{15}
S. Meidt\altaffilmark{1}
E.~J. Murphy\altaffilmark{9} 
N. Rahman\altaffilmark{8}
H.-W. Rix\altaffilmark{1}
H. Roussel\altaffilmark{16}
M. Sauvage\altaffilmark{17}
E. Schinnerer\altaffilmark{1}
R. Skibba\altaffilmark{6}
J.~D. Smith\altaffilmark{4}
C.~D. Wilson\altaffilmark{18}
S. Zibetti\altaffilmark{19}
}

\altaffiltext{1}{Max-Planck-Institut f\"ur Astronomie, K\"onigstuhl 17, D-69117 Heidelberg, Germany    [e-mail: {\em walter@mpia.de}, {\em sandstrom@mpia.de}]}

\altaffiltext{2}{Department of Astrophysical Sciences, Princeton University,
Princeton, NJ 08544, USA}

\altaffiltext{3}{Department of Astronomy, University of Massachusetts, Amherst, MA 01003, USA}

\altaffiltext{4}{Department of Physics and Astronomy, University of
Toledonew, Toledo, OH 43606, USA}

\altaffiltext{5}{Department of Physics \& Astronomy, University of Wyoming,
Laramie, WY 82071, USA}

\altaffiltext{6}{Steward Observatory, University of Arizona, Tucson, AZ 85721, USA}

\altaffiltext{7}{Institute of Astronomy, University of Cambridge, Madingley Road,
Cambridge CB3 0HA, UK}

\altaffiltext{8}{Department of Astronomy, University of Maryland, College Park, MD 20742, USA}

\altaffiltext{9}{Spitzer Science Center, California Institute of Technology, MC
314-6, Pasadena, CA 91125, USA}

\altaffiltext{10}{Leiden Observatory, Leiden University, P.O. Box 9513, 2300 RA
Leiden, The Netherlands}

\altaffiltext{11}{Tianjin Astrophysics Center, Tianjin Normal University, Tianjin
300387, China}

\altaffiltext{12}{NASA Herschel Science Center, IPAC, California Institute of
Technology, Pasadena, CA 91125, USA}

\altaffiltext{13}{INAF - Osservatorio Astrofisico di Arcetri, Largo E. Fermi 5,
50125 Firenze, Italy}

\altaffiltext{14}{Department of Physics and Astronomy, SUNY Stony Brook, Stony
Brook, NY 11794-3800, USA}

\altaffiltext{15}{Hubble Fellow, National Radio Astronomy Observatory, 520 Edgemont Road, Charlottesville, VA 22903, USA}

\altaffiltext{16}{Institut d'Astrophysique de Paris, UMR7095 CNRS, Universit\'e
Pierre \& Marie Curie, 98 bis Boulevard Arago, 75014 Paris, France}

\altaffiltext{17}{CEA/DSM/DAPNIA/Service d'Astrophysique, UMR AIM, CE Saclay, 91191 Gif sur Yvette Cedex}

\altaffiltext{18}{Department of Physics \& Astronomy, McMaster University,
Hamilton, Ontario L8S 4M1, Canada}

\altaffiltext{19}{Dark Cosmology Centre, Niels Bohr Institute, University of Copenhagen, Juliane Maries Vej 30, DK-2100 Copenhagen, Denmark}

\begin{abstract} 

We present the detection of extended ($\sim$30\,kpc$^2$) dust emission
in the tidal \hi\ arm near NGC\,3077 (member of the M\,81 triplet)
using SPIRE on board {\em Herschel}. Dust emission in the tidal arm is
typically detected where the \hi\ column densities are
$>$10$^{21}$\,cm$^{-2}$. The SPIRE band ratios show that the dust in
the tidal arm is significantly colder ($\sim$13\,K) than in NGC\,3077
itself ($\sim$31\,K), consistent with the lower radiation field in the
tidal arm. The total dust mass in the tidal arm is
$\sim$1.8$\times$10$^{6}$\,M$_\odot$ (assuming $\beta$=2),
i.e. substantially larger than the dust mass associated with NGC\,3077
($\sim$2$\times$10$^{5}$\,M$_\odot$). Where dust is detected, the
dust--to--gas ratio is 6$\pm$3$\times$10$^{-3}$, consistent within the
uncertainties with what is found in NGC\,3077 and nearby spiral
galaxies with Galactic metallicities.  The faint \hii\ regions in the
tidal arm can not be responsible for the detected enriched material
and are not the main source of the dust heating in the tidal arm. We
conclude that the interstellar medium (atomic \hi, molecules and dust)
in this tidal feature was pre--enriched and stripped off NGC\,3077
during its recent interaction ($\sim$3$\times$10$^8$\,yr ago) with
M\,82 and M\,81. This implies that interaction can efficiently remove
heavy elements and enriched material (dust, molecular gas) from
galaxies. As interactions were more frequent at large lookback times,
it is conceivable that they could substantially contribute (along with
galactic outflows) to the enrichment of the intergalactic medium.

\end{abstract}

\keywords{galaxies: individual (NGC 3077, Garland) --- ISM: general --- galaxies: ISM --- galaxies: interactions --- infrared: ISM}

\section{Introduction}

It has long been known that tidal interactions can strip off the
extended envelopes of atomic hydrogen (\hi) of interacting
galaxies. This leads to a substantial increase of the \hi\ cross
section (i.e., areal coverage) in such systems. Given these large
cross sections of atomic hydrogen, nearby interacting systems are
thought to be reminiscent of Damped Lyman Alpha (DLA) systems seen in
absorption at high redshift.

One of the most prominent examples of a nearby interacting system is
the M\,81 triplet, a group of galaxies at a distance of $\sim$3.6\,Mpc
(1$'$=1.05\,kpc). Tidal \hi\ features are distributed over
50$\times$100~kpc$^2$ in this system at \hi\ column densities
$>$2$\times$10$^{20}$\,cm$^{-2}$ and interconnect the central three
galaxies M\,81, M\,82 and NGC\,3077 (Yun et al.\ 1994). Early
simulations suggest that the tidal system was created 3$\times$10$^8$
years ago and that the material in the tidal streams eastwards of
NGC\,3077 originally belonged to NGC\,3077 itself (Yun et al.\
1997). The total \hi\ mass of the tidal arm feature around NGC\,3077
is $M$(\hi)=$(3-5)\times10^8$\,M$_\odot$ (van der Hulst 1979, Walter
\& Heithausen 1999, Walter et al.\ 2002a), depending on the exact
integration boundaries. The \hi\ emission in the tidal feature is
kinematically distinct from Galactic cirrus (both in systemic
velocity as well as velocity dispersion, Walter et al.\ 1999, 2002a).

However, rather than being a passive \hi\ tidal feature, this region
has both star--formation and molecular gas, reminiscent of a dwarf
galaxy: As early as 1974, Barbieri et al. detected a `fragmentary
complex of almost stellar objects' in the tidal arm around NGC\,3077,
and subsequent optical observations revealed the presence of a blue
stellar component in this region (dubbed the `Garland', Karachentsev
et al. 1985; Sharina 1991; Sakai \& Madore 2001, Mouhcine \& Ibata
2009). The presence of a young stellar population in the tidal feature
has recently been unambigiously established by Weisz et al.\ (2008)
using deep Hubble Space Telescope ACS observations.

Molecular gas has been detected and mapped in this region through
observations of carbon monoxide (CO, Walter et al.\ 1998, Heithausen
et al.\ 2000). The total star formation rate (SFR) in the tidal
feature is 2.3$\times$10$^{-3}$\,M$_\odot$\,yr$^{-1}$ distributed over
many square-kiloparsec, comparable to what is seen in other low--mass
dwarf galaxies in the M\,81 group of galaxies (Walter, Martin \& Ott
2006). Recent optical spectroscopy of these \hii\ regions indicated
that their metallicity is much higher than expected based on the total
blue magnitude of the stellar system in the tidal arm, possibly
reaching a value as high as found in NGC\,3077 itself, i.e. close to
the Galactic one (Croxall et al.\ 2009). Here we report the detection
of substantial amounts of cold dust on kpc scales in the tidal feature
around NGC\,3077 based on {\em Herschel} SPIRE observations.  

\begin{figure*}
\centering
\includegraphics[width=14cm,angle=0]{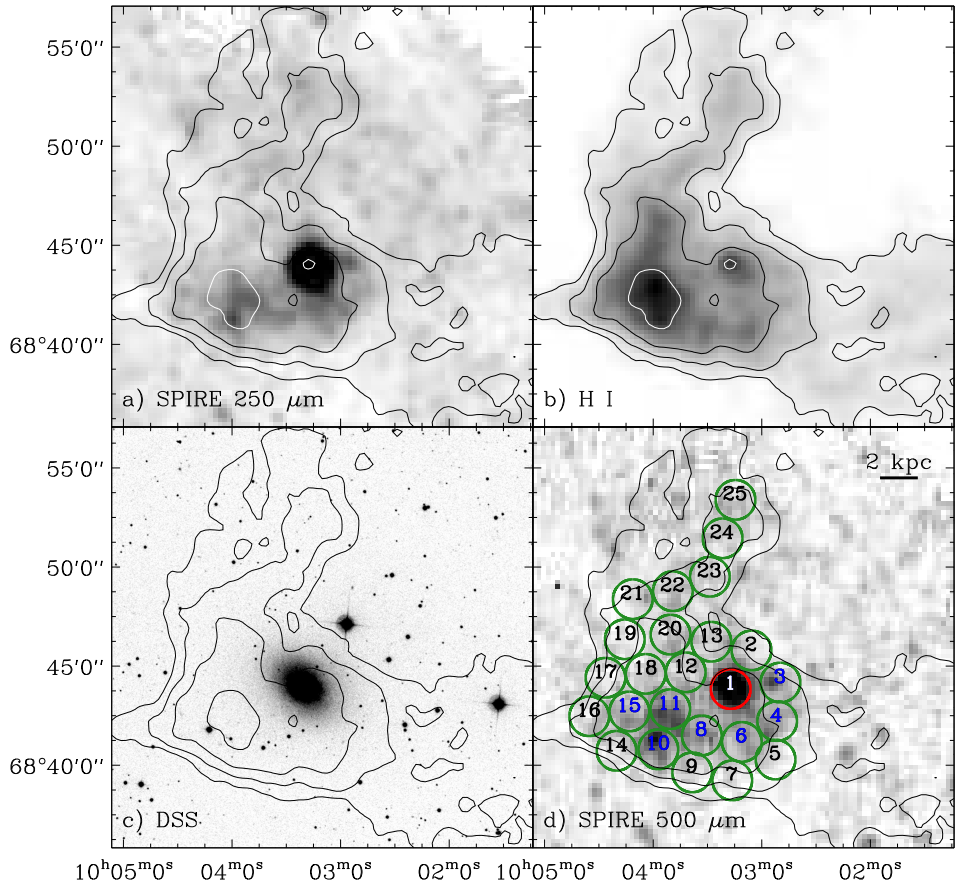}
\vspace{-0.5cm}
\caption{Multi--wavelength view of NGC\,3077 and the surrounding tidal
\hi\ system (North is top, East is left). {\em Top left:} SPIRE 250$\mu$m image of NGC\,3077 and
its surrounding. The contours are the \hi\ intensities shown in the top
right panel. {\em Top right:} Distribution of the tidal \hi\ complex
around NGC\,3077 at 40$^{\prime\prime}$ resolution (greyscale). Contours are shown at \hi\
columns of 2, 5, 10 and 20$\times$10$^{20}$\,cm$^{-2}$ at that
resolution (from Walter et al.\ 2008). {\em Bottom left:} DSS image
of NGC\,3077 and its surroundings with the \hi\ contours superimposed. {\em
Bottom right:} SPIRE 500$\mu$m image, with \hi\  contours and apertures
superimposed. Aperture \#1 is centred on NGC\,3077 itself.
}
\end{figure*}

\section{Observations}

NGC\,3077 was observed with the SPIRE instrument on {\em Herschel}
(Pilbratt et al. 2010) in `scan map' mode as part of the Open Time Key
Project KINGFISH (PI: R.~Kennicutt) on March 28th, 2010. The galaxy
and its surroundings were homogeneously covered using 11~arcmin
scan--legs (AOR length: 0.65 hours).  Data were reduced using the
standard calibration products and algorithms available in HIPE (the
{\em Herschel} Interactive Processing Environment), version 2.0.0.
The determination of the background in NGC\,3077 is complicated by the
extended \hi\ filaments around the region of interest. The background
was defined away from the \hi\ tidal tails (here defined as column
densities $>$2$\times$10$^{20}$\,cm$^{-2}$, including NGC\,3077
itself), to avoid subtracting real dust emission that may be
coincident with \hi\ features. The SPIRE data were calibrated in units
of Jy\,beam$^{-1}$, which were converted to MJy\,sr$^{-1}$ by assuming
beam sizes of 501, 944, and 1924 square arcseconds (FWHM:
18$^{\prime\prime}$, 25$^{\prime\prime}$ and 37$^{\prime\prime}$) for
the 250, 350, and 500 $\mu$m bands. For the quantitative analysis we
convolved all data to the SPIRE 500 $\mu$m resolution using the
kernels described by Gordon et al. (2008) and updated by those authors
for {\em Herschel}. We have also applied calibration correction
factors of (1.02, 1.05, 0.94) at (250, 350, 500) $\mu$m to the values
as given in the observers manual. We selected 44 background apertures
(radius: 1$^{\prime}$) away from NGC\,3077 and the tidal \hi\
arms. The respective flux histograms of these apertures at 250, 350
and 500 $\mu$m are centered on zero, as expected given the background
subtraction procedure, and have 1\,$\sigma$ widths of (0.33, 0.16 and
0.086)~MJy\,sr$^{-1}$ at (250, 350, and 500) $\mu$m which we adopt as
the uncertainty in the tidal feature aperture measurements (we do not
include the 15\% SPIRE calibration uncertainty which will affect all
three bands in a similar way).  \\[0.5cm]

\section{Results}

\subsection{Dust in the Tidal \hi\  Arm}

Figure~1 presents a multi--wavelength view of the tidal features
around NGC\,3077. The top left panel shows the SPIRE map at 250$\mu$m
(convolved to 40$^{\prime\prime}$ resolution, i.e. approximately the
500$\mu$m beam). The brightest source in the field is NGC\,3077
itself, but significant emission is detected east of the main body of
the galaxy. The nuclear starburst in NGC\,3077 is distributed over
scales of $\sim$10$^{\prime\prime}$ (Martin et al.\ 1997, Ott et al.\
2003, Walter et al.\ 2006), which is unresolved at SPIRE's
resolution. The extended dust emission towards the east is spatially
coincident with \hi\ column densities $\geq$10$^{21}$\,cm$^{-2}$ in
the tidal feature (Fig.~1, top right; \hi\ data are taken from THINGS,
`The \hi\ Nearby Galaxy Survey', Walter et al.\ 2008). The apparent
correlation of \hi\ emission with dust seen in the SPIRE data (in
particular towards high \hi\ column densities) implies that the dust
emission is indeed physically coincident with the tidal arm and does
not arise from Galactic cirrus which is observed towards the direction
of the M\,81 group (e.g. Davies et al.\ 2010, Sollima et al.\ 2010).

\subsection{Selection of Apertures}

To quantify the dust properties, we have selected 25 apertures of
radius 1$'$ ($\sim$1\,kpc) each from visual inspection of the SPIRE
250\,$\mu$m map. The locations (and numbers) of the apertures are
shown in the bottom right panel of Fig.~1 (green circles) and the
positions are tabulated in Tab.~1. The red circle indicates the
aperture (\#1) centered on NGC\,3077. For each aperture, we have
derived the flux in the SPIRE bands (Tab.~1). The error in this
measurement is derived from the dispersion of 44 background apertures
of the same size that are located away from the \hi\ tidal arms (at
column densities $<$2$\times$10$^{20}$\,cm$^{-2}$). This dispersion is
mainly driven by unresolved background sources at high redshift. The
\hi\ column densities (at 40$^{\prime\prime}$ resolution) for each
aperture are also given in Tab.~1 --- all column densities are
$>$4\,M$_\odot$\,pc$^{-2}$
(5$\times$10$^{20}$\,cm$^{-2}$). Significant dust emission (here
defined as a $>$3$\sigma$ detection at 250\,$\mu$m) is detected in 8
apertures, i.e. an area of $\sim$30\,kpc$^{2}$. These apertures are
labeled in blue in Fig.~1 (bottom right) and marked with a boldface
font in Tab.~1. Averaging the other apertures also yields a
statistical detection of dust emission.

\begin{deluxetable*}{rllrrrlrr}
\tablecaption{Aperture Measurements}
\tablewidth{0pt}
\tablehead{
\colhead{\#\tablenotemark{a}} & \colhead{RA} & \colhead{DEC} & \colhead{F(250$\mu$m)\tablenotemark{b}}  & \colhead{F(350$\mu$m)\tablenotemark{b}} & \colhead{F(500$\mu$m)\tablenotemark{b}}  & \colhead{$\Sigma_{\rm{dust}\tablenotemark{c}}$} & \colhead{$\Sigma_{\rm \hi}$\tablenotemark{d}} & \colhead{$\Sigma_{\rm SFR}$\tablenotemark{e}}\\
\colhead{}   & \colhead{J2000.0}& \colhead{J2000.0}    & \colhead{MJy\,sr$^{-1}$}& \colhead{MJy\,sr$^{-1}$}&\colhead{MJy\,sr$^{-1}$}  & \colhead{M$_\odot$\,pc$^{-2}$} & \colhead{M$_\odot$\,pc$^{-2}$} & \colhead{10$^{-4}$\,M$_\odot$\,yr$^{-1}$\,kpc$^{-2}$}
}
\startdata
{\bf 1} & 10 03 17.18 & +68 43 53.0 & 23.61 &  8.41 & 3.07 &  0.054$\pm$0.005 &  12.42 & 220 \\
 2 & 10 03 05.57 & +68 45 49.4 &  0.73 &  0.34 & 0.15 &       \nodata         &   2.95 &\nodata \\
{\bf 3} & 10 02 49.53 & +68 44 12.2 &  1.54 &  0.91 & 0.48 &  0.053$\pm$0.015 &   4.88 &\nodata \\
{\bf 4} & 10 02 51.35 & +68 42 15.7 &  1.71 &  0.99 & 0.53 &  0.058$\pm$0.016 &   7.04 &\nodata \\
 5 & 10 02 52.28 & +68 40 19.3 &  0.02 &  0.11 & 0.09 &       \nodata         &   6.04 &\nodata \\
{\bf 6} & 10 03 10.96 & +68 41 12.8 &  2.63 &  1.32 & 0.58 &  0.075$\pm$0.020 &  10.18 &\nodata \\
 7 & 10 03 16.30 & +68 39 16.3 & -0.10 &  0.05 & 0.05 &       \nodata         &    5.63&\nodata \\
{\bf 8} & 10 03 33.21 & +68 41 32.2 &  2.53 &  1.35 & 0.65 &  0.078$\pm$0.021 &  10.69 & 0.65 \\
 9 & 10 03 38.53 & +68 39 40.5 &  0.29 &  0.31 & 0.17 &       \nodata         &   7.69 &\nodata \\
{\bf 10} & 10 03 57.30 & +68 40 50.1 &  2.74 &  1.73 & 0.95 & 0.101$\pm$0.027 &  14.78 & 4.00 \\
{\bf 11} & 10 03 50.72 & +68 42 51.7 &  2.65 &  1.68 & 0.91 & 0.097$\pm$0.026 &  16.23 & 1.23 \\
12 & 10 03 42.18 & +68 44 43.2 &  0.76 &  0.57 & 0.34 &       \nodata         &   8.16 &\nodata \\
13 & 10 03 27.96 & +68 46 15.9 &  0.84 &  0.51 & 0.31 &       \nodata         &   3.30 &\nodata \\
14 & 10 04 20.44 & +68 40 45.2 &  0.63 &  0.41 & 0.21 &       \nodata         &   6.39 &\nodata \\
{\bf 15} & 10 04 13.43 & +68 42 44.4 &  1.54 &  1.02 & 0.52 &  0.057$\pm$0.016 &  15.62& 0.29\\
16 & 10 04 35.72 & +68 42 29.4 &  0.59 &  0.41 & 0.21       & \nodata &   6.30 & 1.09\\
17 & 10 04 26.79 & +68 44 26.1 &  0.19 &  0.16 & 0.10       & \nodata &   5.23 &\nodata \\
18 & 10 04 04.49 & +68 44 38.4 &  0.63 &  0.42 & 0.26       & \nodata &  12.60 &\nodata \\
19 & 10 04 16.16 & +68 46 22.8 & -0.23 & -0.07 & 0.01       & \nodata &   6.49 &\nodata \\
20 & 10 03 50.70 & +68 46 37.3 &  0.54 &  0.43 & 0.31       & \nodata &   8.93 &\nodata \\
21 & 10 04 11.77 & +68 48 24.2 &  0.04 &  0.11 & 0.13       & \nodata &   4.71 &\nodata \\
22 & 10 03 49.41 & +68 48 48.8 &  0.89 &  0.51 & 0.29       & \nodata &   5.94 &\nodata \\
23 & 10 03 28.82 & +68 49 32.7 &  0.72 &  0.39 & 0.19       & \nodata &   4.58 &\nodata \\
24 & 10 03 21.66 & +68 51 29.2 &  0.78 &  0.42 & 0.23       & \nodata &   5.35 &\nodata \\
25 & 10 03 14.47 & +68 53 25.6 &  0.78 &  0.38 & 0.24       & \nodata &   4.36 &\nodata 
\enddata
\tablenotetext{a}{Aperture number in {\bf bold} indicate regions with significant emission (i.e. $>3$\,$\sigma$ at 250\,$\mu$m). Dust mass densities have been derived for those regions assuming a temperature of T=12.6~K (only exception: aperture \#1 [NGC\,3077, T=30.6\,K]) and $\beta$=2 (see Sec.~3.4).}
\tablenotetext{b}{The uncertainties in the SPIRE measurements are (0.33, 0.16 and 0.086)~MJy\,sr$^{-1}$ at (250, 350, and 500) $\mu$m, respectively, based on background aperture measurement. These uncertainties do not include the SPIRE calibration uncertainties of $\sim$15\%.} 
\tablenotetext{c}{The uncertainty in dust mass surface densities have been derived using different temperatures ($\pm$1\,K) as the fitting errors give unrealistically small numbers.} 
\tablenotetext{d}{\hi\ surface density uncertainties are typically $<$10\%.}
\tablenotetext{e}{SFR surface densities are derived from the values given in Walter et al.\ 2006 (tidal arm) and Kennicutt et al.\ 2008 (NGC\,3077).}
\end{deluxetable*}

\subsection{SPIRE Dust SED}

Figure~2 shows the SPIRE measurements for NGC\,3077 (aperture \#1, red
curve) and the sum of all
detected tidal regions (green curve, boldface aperture numbers in
Table~1). A comparison of the two SEDs directly shows that the
250$\mu$m/500$\mu$m ratio is significantly lower in the tidal feature
compared to the main body of NGC\,3077, reflecting a lower temperature
in the tidal region.  Given the lower temperature in the tidal arm and
the comparable total fluxes in the SPIRE bands this immediately
implies that the dust mass in the tidal feature is larger than in
NGC\,3077 itself (see more detailed discussion in Sec.~3.6).

We have attempted to use Spitzer MIPS 160$\mu$m imaging to further
constrain the SED towards shorter wavelengths. The main body of
NGC\,3077 is very bright, and we plot the corresponding flux density
in Fig.~2.  Because of coverage and extended cirrus emission, the
background of the MIPS 160$\mu$m is however badly behaved at the flux
levels of interest, leading to significant error bars in the flux
determination. We nevertheless derived a 160$\mu$m flux density for
the brightest aperture (\#10, the error bar is dominated by the
background uncertainties) --- the complete SED for this aperture is
also plotted in Fig.~2 for comparison.

\subsection{Dust Temperature}

We have used a blackbody fit and the Li \& Draine (2001) dust
emissivity (equivalent to a modified blackbody with $\beta$=2) to
derive temperatures and dust masses. The temperature for NGC\,3077
(aperture \#1) is 30.6$\pm$2\,K and is in agreement with high
temperatures expected for starbursts and the central temperatures
derived for a sub--sample of the KINGFISH galaxies presented by
Engelbracht et al.\ (2010) of T$_{\rm
center}$=25.7$\pm$1\,K. The
temperature in the tidal feature is much lower. Based on the SPIRE
data only, we derive an average temperature of
T=12.0$\pm$1\,K. In the brightest aperture of the tidal
feature (\#10) the addition of the 160$\mu$m flux estimate increases
the temperature slightly (12.6\,K), but is consistent with the
`SPIRE--only' value. Changing $\beta$ to a value of 1.5 increases the
temperature by $\sim$3\,K (see dashed SED fits in Fig.~2). In the
following we will adopt $\beta$=2 and T=12.6$\pm$2 for the tidal
feature apertures but note that the temperature (and the corresponding
masses) depend on the exact choice of dust model.  

\subsection{Radiation Field}

Finding low dust temperatures may not be unexpected as the intensity
of the radiation field in the tidal arm is presumably low: the total
star formation in the entire tidal feature is only
2.3$\times$10$^{-3}$\,M$_\odot$\,yr$^{-1}$ (based on H$\alpha$
observations, Walter et al. 2006). This SFR is consistent with the
rate determined in Weisz et al.\ (2008) through stellar population
studies, and GALEX measurements that cover regions \#10 and \#11,
implying that the SFR in the tidal feature was roughly constant over
the recent past (few hundred million years).  For those apertures that
include \hii\ regions we sum up their H$\alpha$ luminosities using the
values given in Walter et al.\ (2006) and give the star formation rate
surface densities (averaged over the size of our apertures) in
Tab.~1. We plot the 250$\mu$m/500$\mu$m ratios for these aperatures as
a function of the H$\alpha$ luminosity surface density in
Figure~3. Star formation rates can be estimated using
SFR[M$_\odot$\,yr$^{-1}$]=7.9$\times$10$^{-42}$\,L(H$\alpha$)[erg\,s$^{-1}$]
(Kennicutt 1998, see top x-axis in Fig.~3; choosing a Kroupa IMF would
decrease the SFRs by $\sim$30\%) and the brightest region outside of
NGC\,3077 (aperture \#10) accounts for $\sim$50\% of the total SFR in
the tidal complex (1.3$\times$10$^{-3}$\,M$_\odot$\,yr$^{-1}$, or a
SFR surface density for this aperture of
4$\times$10$^{-4}$\,M$_\odot$\,yr$^{-1}$\,kpc$^{-2}$).  For reference,
single massive stars create \hii\ regions with H$\alpha$ luminosities
of L$_{\rm H\alpha, O7}\approx 5\times10^{36}$\,erg\,s$^{-1}$ and
L$_{\rm H\alpha, O5}\approx 5\times10^{37}$\,erg\,s$^{-1}$ (e.g.,
Devereux et al.\ 1997), i.e. the total H$\alpha$ flux in the tidal
feature can in principle be created by a few massive stars.  Figure~3
shows that the dust temperature does not change as a function of the
star formation rate surface density over two orders of magnitude of
the latter. This suggests that the currently observed star formation
does not provide the main heating source for the dust on kpc scales
and we speculate that other mechanisms may be responsible for the dust
heating.

Assuming $G_0\sim$4$\times$10$^{-3}$ for the intergalactic radiation
field (Sternberg et al.\ 2002, note that $G_0$=1 for the Galactic
interstellar radiation field), that the temperature of dust goes as
$T\sim G_0^{1/(4+\beta)}$ and $\beta$=2, and scaling from the results
by Li \& Draine (2001) for the Galactic interstellar radiation field
(16\,K for a silicate grain, 20\,K for a graphite grain, Fig~3 in Li
\& Draine 2001), we estimate that the dust in the tidal feature should
have a temperature of 6-8\,K in the intergalactic field. This suggests
that other heating sources (the radiation field of NGC 3077, an older
stellar population in the feature as seen in the HST imaging by Weisz
et al.\ 2008 and/or shocks) must be partly responsible for heating the
dust to our derived temperature of 12$\pm$2 K. The 250$\mu$m/500$\mu$m
ratio does not change as a function of projected distance to NGC\,3077
within our uncertainties but we note that the measurements only span a
small range in distances.

\begin{figure}
\centering
\includegraphics[width=9cm,angle=0]{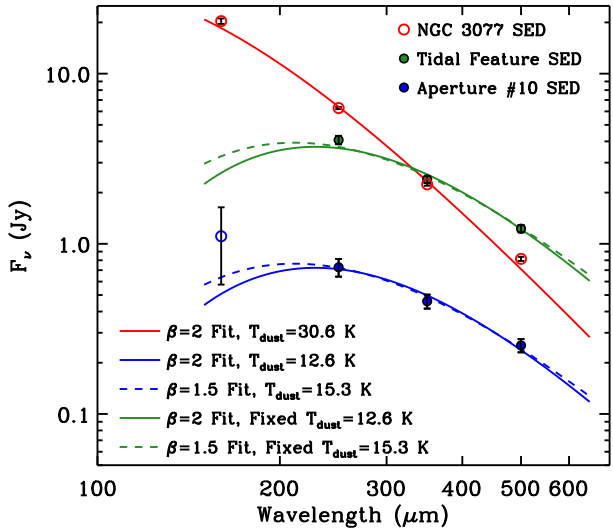}

\caption{SPIRE dust SEDs for NGC\,3077 (aperture \#1, red color), the
entire tidal filament (summing up apertures that contain significant
emission, green) and the brightest aperture (\#10, blue). The full
lines correspond to fits assuming $\beta$=2, the dashed curves
indicate $\beta=1.5$ (fitted temperatures are given in the
Figure). The SPIRE calibration uncertainties ($\sim$15\%) are not
included in the error bars.}

\end{figure}

\begin{figure}
\centering
\includegraphics[width=9cm,angle=0]{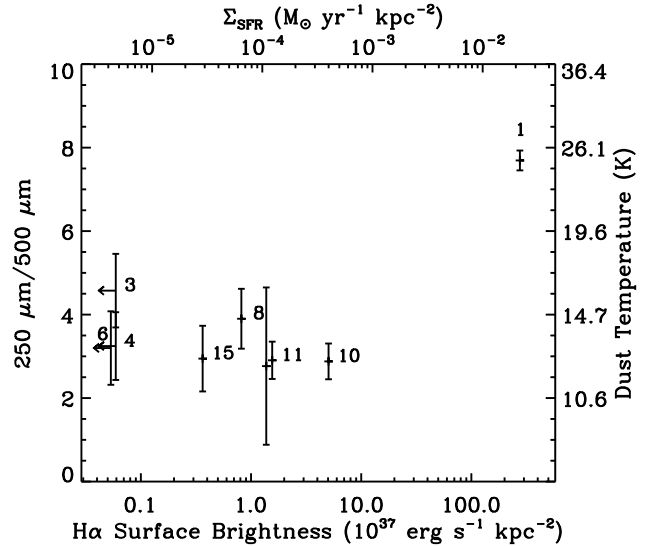}
\caption{250$\mu$m/500$\mu$m ratio in the tidal feature as a function
of H$\alpha$ luminosity in a given aperture (numbers indicate aperture
numbers). The temperatures given on the right y--axis are based on
this 250$\mu$m/500$\mu$m ratio and assume a blackbody with $\beta$=2.}
\end{figure}

\subsection{Dust Masses and Dust/Gas Ratios}

From the SED fits discussed above we derive dust mass surface
densities for each of our apertures (Tab.~1) with an average value of
7.4$\times$10$^{-2}$\,M$_\odot$\,pc$^{-2}$ ($\beta$=2, T=12.6\,K). We
derive a total dust mass in the tidal feature (excluding NGC\,3077 and
upper limits) of $\sim$1.8$\pm$0.9$\times$10$^6$\,M$_\odot$ (the large
error bar includes uncertainties in dust properties, see below). Our
simple dust mass estimate is consistent with results from more
detailed modeling using updated Draine \& Li (2007) models (Aniano et
al., in prep.). The dust mass of NGC\,3077 is only
1.9$\times$10$^5$\,M$_\odot$ ($\beta$=2, T=30.6\,K), in reasonable
agreement with the mass estimates by Price \& Gullixson (1989) of the
absorbing central dust clouds based on NIR observations (they report a
mass of $\sim$10$^5$\,M$_\odot$). Both mass estimates depend on the
choice of $\beta$ and the temperature. For example, if we chose $\beta$=1.5
for the tidal feature, the dust mass would go down by a factor of
$\sim$2 (as the temperature would increase by $\sim$3\,K,
Sec.~3.4). Likewise, lowering the dust temperature in NGC\,3077 by
5\,K would increase its dust mass by a factor of $\sim$2. Given these
large systematic uncertainties involved we assign a conservative error
of 50\% to the dust mass of the tidal feature. 

For the average \hi\ surface mass densities of the apertures where
dust emission was significantly detected, we derive an average value of
11.3\,M$_\odot$\,pc$^{-2}$ (Tab.~1), i.e. more than two orders of
magnitude larger than the dust surface densitites. The average
dust--to--\hi\ gas ratio is 6.5$\pm$3$\times$10$^{-3}$, and the ratio
does not vary within the errors between apertures and is not a
function of \hi\ surface density (the corresponding value for
NGC\,3077 (aperture \#1) is 0.0043). Within the uncertainties this
value is close to the average dust--to--gas ratio of M$_{\rm
dust}$/M$_{\rm \hi}\approx0.005$ derived by Draine et al.\ (2007) for
12 spiral galaxies of the SINGS sample for which Spitzer and SCUBA
measurements were available. As noted by Draine et al.\ (2007), a
M$_{\rm dust}$/M$_{\rm \hi}$ ratio between 0.003 and 0.01 is
consistent with galactic metallicities between 0.3 and 1 times solar,
if a similar fraction of heavy elements is in the form of dust as in
the Milky Way.  Our derived dust--to--gas ratio thus indicates that
the tidal arm is substantially chemically enriched, in agreement with
the \hii\ region metallicities derived by Croxall et al.\ (2009). For
comparison, the M$_{\rm dust}$/M$_{\rm HI}$ ratio for metal--poor
dwarf irregular galaxies in the (more extended) M\,81 group of
galaxies is up to one order of magnitude lower than found here
(Walter et al.\ 2007).

So far, we have only considered the \hi\ gas. Molecular gas has been
mapped in the CO(1--0) transition in the tidal feature over a
restricted area (Walter et al.\ 1998, Heithausen \& Walter 2000,
Walter et al.\ 2006). The two main molecular complexes (region number
\#1 and \#2 in Heithausen \& Walter 2000) are spatially coincident
with aperture numbers~10 and~11 in this study. The implied molecular
gas masses are dependent on the choice of the H$_2$--to--CO conversion
factor X$_{\rm CO}$. Heithausen \& Walter (2000) used a conversion
factor of X$_{\rm
CO}$=8$\times$10$^{20}$cm$^{-2}$(K\,km\,s$^{-1}$)$^{-1}$, a factor of
$\sim$4 larger than the Galactic conversion factor. At the time, their
choice was driven by the assumption that the tidal region was
metal--poor. Given the abundance of dust and dust--to--gas ratios
similar to nearby spiral galaxies, as well as the \hii\ region
metallicity measurements (indicating metallicities as high as
Galactic, Croxall et al.\ 2009), it now seems to be more appropriate
to use the Galactic conversion factor in this tidal feature. This
implies that the masses given in Heithausen \& Walter (2000) should be
divided by a factor of 4. We thus adopt molecular gas masses of
$\sim$1$\times$10$^{6}$\,M$_\odot$ and
$\sim$4$\times$10$^{6}$\,M$_\odot$ for our apertures \# 10 and 11,
respectively (comparable to the H$_2$ mass present in the centre of
NGC\,3077, Meier et al.\ 2001, Walter et al.\ 2002a). Averaged over
our r=1\,kpc--sized apertures this corresponds to a molecular surface
density between 1.1--0.3\,M$_\odot$\,pc$^{-2}$ (i.e., the ISM in the
tidal feature is dominated by the atomic gas phase even in the
CO--brightest regions.

\section{Concluding Remarks}

The detection of significant amount of dust
($\sim$1.8$\pm$0.9$\times$10$^{6}$\,M$_\odot$) in the tidal feature
near NGC\,3077, distributed over 30~square--kiloparsecs, which is
significantly larger than the dust mass present in the parent galaxy
(irrespective of the choice of $\beta$ and reasonable temperatures),
raises questions on the origin of the enriched material. The ongoing
star formation could potentially enrich the medium in the tidal
feature. If we assumed a constant star formation rate since the
creation of the feature ($3\times10^8$\,yr ago, Yun et al.\ 1994) we
estimate a total mass of newly formed stars of
$\sim7\times10^{5}$\,M$_{\odot}$, i.e. less than the total amount of
dust that is present. This implies that the current rate of star
formation activity can not have created the present dust. Also, the
chemical enrichment due to the \hii\ regions in the tidal arm is
expected to lead to a metallicity of only Z$\sim$0.002 Solar (Walter
et al.\ 2006) over the last $3\times10^8$\,yr, i.e., much lower than
what is measured by Croxall et al.\ (2009) and implied by our
dust--to--gas ratio. We conclude that the tidal arm material was
pre--enriched, and likely belonged to NGC\,3077 (which has a
metallicity similar to the tidal \hii\ regions, Croxall et al.\ 2009)
before the interaction.

Our findings imply that interactions between galaxies can efficiently
remove heavy elements, dust and molecules from a galaxy. In the case
discussed here, significantly more dust mass is found in the tidal arm
than in the parent galaxy NGC\,3077 (the same holds true for the
atomic and molecular gas). Interactions thus appear to have the
potential to alter the chemical evolution of a galaxy dramatically and
to expel an interstellar medium that is enriched by heavy elements. As
interactions have been more frequent at larger lookback times it is
conceivable that this mechanism can (in addition to outflows, e.g.,
Steidel et al.\ 2010) effectively enrich the intergalactic medium (see
also Roussel et al.\ 2010, Walter et al.\ 2002b). The tidal system
discussed here would be classified as a damped Lyman--alpha absorber
(DLA) with a cross section of roughly 30 kpc$^2$ (e.g. Wolfe, Gawiser
\& Prochaska 2005, Zwaan et al.\ 2008).

\acknowledgements FW acknowledges the hospitality of the Aspen Center
for Physics. {\em
Herschel} is an ESA space observatory with science instruments
provided by European-led Principal Investigator consortia and with
important participation from NASA.

\end{document}